\documentclass[12pt]{article}
\usepackage{epsfig,amssymb,euscript,bbold}
\newcommand{\be}{\begin{equation}}
\newcommand{\ee}{\end{equation}}
\newcommand{\bea}{\begin{eqnarray}}
\newcommand{\eea}{\end{eqnarray}}
\newcommand{\ba}{\begin{array}}
\newcommand{\p}[1]{(\ref{#1})}
\newcommand{\ea}{\end{array}}

\def\bbox{{\,\lower0.9pt\vbox{\hrule \hbox{\vrule height 0.2 cm
\hskip 0.2 cm \vrule height 0.2 cm}\hrule}\,}}
\newcommand{\dsl}{\pa \kern-0.5em /}

\newcommand{\nn}{\nonumber \\}

\newcommand{\diag}{{\rm diag}\,}

\newcommand{\dcy}{d}               
\newcommand{\newa}{a}               
\newcommand{\newf}{h}               

\def\ds{\raise.15ex\hbox{/}\kern-.57em\partial}
\def\Ds{\,\raise.15ex\hbox{/}\mkern-13.5mu D}
\font\mybb=msbm10 at 10pt
\def\bb#1{\hbox{\mybb#1}}

\def\bR {\bb{R}}
\def\bT {T}
\def\bH {H}
\def\bC {\bb{C}}
\def\bS {S}
\def\bP {\bb{P}}
\def\bK {K}

\begin{document}

\makeatletter
\renewcommand{\theequation}{\thesection.\arabic{equation}}
\@addtoreset{equation}{section}
\makeatother

\baselineskip 18pt


\begin{titlepage}
\vfill
\begin{flushright}
QMW-PH-00-16\\
hep-th/0012195\\
\end{flushright}

\vfill

\begin{center}
\baselineskip=16pt
{\Large\bf M-Fivebranes Wrapped on} 
\\
{\Large\bf Supersymmetric Cycles} 
\vskip 10.mm
{Jerome P. Gauntlett$^{1}$, ~Nakwoo Kim$^{2}$ and Daniel Waldram$^{3}$}\\
\vskip 1cm
{\small\it
Department of Physics\\
Queen Mary, University of London\\
Mile End Rd, London E1 4NS, UK}\\
\vspace{6pt}
\end{center}
\vfill
\par
\begin{center}
{\bf ABSTRACT}
\end{center}
\begin{quote}
We construct supergravity solutions dual to
the twisted field theories arising when
M-theory fivebranes wrap general supersymmetric cycles.
The solutions are constructed in maximal $D=7$ gauged supergravity
and then uplifted to $D=11$. Our analysis covers 
K\"{a}hler, special Lagrangian and exceptional calibrated cycles. 
The metrics on the cycles are Einstein, but do
not necessarily have constant curvature. 
We find many new examples of AdS/CFT duality, corresponding 
to the IR superconformal fixed points of the 
twisted field theories.
\vfill
\vskip 5mm
\hrule width 5.cm
\vskip 5mm
{\small
\noindent $^1$ E-mail: j.p.gauntlett@qmw.ac.uk \\
\noindent $^2$ E-mail: n.kim@qmw.ac.uk \\
\noindent $^3$ E-mail: d.j.waldram@qmw.ac.uk \\
}
\end{quote}
\end{titlepage}
\setcounter{equation}{0}

\section{Introduction}

The supergravity duals of the twisted field theories arising
on branes wrapping supersymmetric cycles \cite{bvs} have recently 
been investigated in \cite{malnun,malnuntwo,agk,no} (for related 
solutions see \cite{AO,CV,FS1,FS2,FS3}). 
The cases that have been considered include
fivebranes and D3-branes wrapping holomorphic 
curves \cite{malnun,malnuntwo}, and fivebranes \cite{agk} and
D3-branes \cite{no} wrapping associative
three-cycles. Here we will extend these investigations
by analysing M-fivebranes wrapping all other supersymmetric cycles.
The new cases covered here are K\"{a}hler four-cycles, 
special Lagrangian three- four- and five-cycles, 
co-associative four-cycles and Cayley four-cycles.

As in previous work it will be convenient to first construct the 
solutions in $D=7$ gauged supergravity and then uplift to obtain
the $D=11$ solutions. Rather than working with different truncated
versions of gauged supergravity we will present a unified
treatment by working directly with the maximal $SO(5)$ gauged supergravity of
\cite{vn}. We then employ the results of \cite{vntwo,vnthree} to uplift to
$D=11$. This approach has the virtue of highlighting the universal 
aspects of the various supergravity solutions. 

An ingredient in the supergravity solutions will be a metric on the
supersymmetric cycle $\Sigma_d$. This metric is required to be Einstein,
satisfying $R_{ij}=l g_{ij}$, where, factoring out the overall scale
of $\Sigma_d$, we have $l=0,\pm 1$. The metric on the
special Lagrangian cycles will be further restricted to
have constant curvature. For the exceptional four-cycles we
impose that the metrics are half-conformally flat 
i.e., the Weyl tensor is self-dual. For the K\"{a}hler cycles it is
sufficient that the metric is K\"{a}hler-Einstein. 
Setting $l=-1$, for all cases except for
K\"{a}hler four-cycles in Calabi-Yau three-folds, we find explicit solutions
of the form $AdS_{7-d}\times\Sigma_d$. These solutions are the gravity
duals of the superconformal theories arising on the wrapped brane. 
For the single case of SLAG five-cycles we also find an exact solution 
with $l=1$ of the form $AdS_2\times \bS^5$.

We begin in section 2 by analysing the general aspects of
the BPS equations arising from $D=7$ gauged supergravity. 
This is followed in sections 3--5 with a discussion of the 
BPS equations for the different cases as well as a presentation
of the $AdS_{7-d}\times \Sigma_d$ solutions and the formulae to uplift
to $D=11$.  Section 6 contains some numerical analysis of the
BPS equations where we demonstrate the flows when $l=-1$ 
from an $AdS_7$ type regions to the $AdS_{7-d}\times\Sigma_d$ solutions.
We also analyse the BPS equations and the singularities of the general 
flows with $l=\pm 1$. Section 7 briefly concludes.

\section{Maximal $D=7$ Gauged Supergravity}

The Lagrangian for the bosonic fields of 
maximal gauged supergravity in $D=7$  
is given by \cite{vn}
\bea
&&2{\cal L}=e\left[R+\frac{1}{2}m^2(T^2-2T_{ij}T^{ij})-
P_{\mu ij}P^{\mu ij}-\frac{1}{2}({\Pi_A}^i{\Pi_B}^j
F_{\mu\nu}^{AB})^2-m^2({{\Pi^{-1}}_i}^AS_{\mu\nu\rho,A})^2\right]\nn
&&-{6m}\delta^{AB}S_A\wedge F_B +{\sqrt 3}
\epsilon_{ABCDE}\delta^{AG}S_G\wedge F^{BC}\wedge
F^{DE}
+{1\over 8m}(2\Omega_5[B]-\Omega_3[B])
\eea                  
Here $A,B=1,\dots,5$ denote indices of the $SO(5)_g$ gauge-group, while
$i,j=1,\dots, 5$ denote indices of the $SO(5)_c$ local composite
gauge-group, which are raised and lowered with $\delta^{ij}$ and
$\delta_{ij}$. The 14 scalar fields ${\Pi_A}^i$ are given by the coset
$SL(5,\bR)/SO(5)_c$ and transform as a ${\bf 5}$ under both $SO(5)_g$
(from the left) and $SO(5)_c$ (from the right). The term that gives
the scalar kinetic term, $P_{\mu ij}$, and the $SO(5)_c$ composite
gauge-field, $Q_{\mu ij}$, are defined as the symmetric and
antisymmetric parts of ${(\Pi^{-1})_i}^A\left(\delta_A{}^B \partial_\mu + 
g B_{\mu\,A}{}^B\right){\Pi_B}^k \delta_{kj}$, respectively. 
Here $B_A{}^B$ are the $SO(5)_g$ gauge-fields with field strength 
$F^{AB}=\delta^{AC}{F_C}^B$, and note that the gauge coupling constant
is given by $g=2m$. The four-form field strength $F_A$ for the three-form
$S_A$, is the covariant derivative $F_A=dS_A+gB_A\;^BS_B$. The
potential terms for the scalar fields are expressed in terms of
$T_{ij}={{\Pi^{-1}}_i}^A{{\Pi^{-1}}_j}^B\delta_{AB}$ with
$T=\delta^{ij}T_{ij}$. Finally, $\Omega_3[B]$ and  $\Omega_5[B]$ are
Chern-Simons forms for the gauge-fields $B$ that will not play a role
in this paper.  

The supersymmetry transformations of the fermions
are given by
\bea\label{susytran}
\delta\psi_{\mu}&=&\nabla_{\mu}\epsilon+\frac{1}{20}mT
\gamma_{\mu}\epsilon -\frac{1}{40}(\gamma_{\mu}{}^{\nu\rho}-8
\delta_{\mu}{}^{\nu}\gamma^{\rho})\Gamma_{ij}\epsilon\, {\Pi_A}^i{\Pi_B}^j
F_{\nu\rho}^{AB}          
\nonumber\\
&& + \frac{m}{10\sqrt{3}}(\gamma_{\mu}{}^{\nu\rho\sigma}
-\frac{9}{2}\delta_{\mu}{}^{\nu}\gamma^{\rho\sigma})\Gamma^i\epsilon\,
{{\Pi^{-1}}_i}^AS_{\nu\rho\sigma,A} \nn
\delta\lambda_i&=&
   \frac{1}{2}\gamma^{\mu}\Gamma^j \epsilon\, P_{\mu ij}
   + \frac{1}{2}m(T_{ij}-\frac{1}{5}T\delta_{ij})\Gamma^j\epsilon
   + \frac{1}{16}\gamma^{\mu\nu}(\Gamma_{kl}\Gamma_i
       -\frac{1}{5}\Gamma_i\Gamma_{kl})
       \epsilon\, {\Pi_A}^k{\Pi_B}^l F_{\mu\nu}^{AB} \nonumber\\
   && + \frac{m}{20\sqrt{3}}\gamma^{\mu\nu\rho}(\Gamma_i{}^j-4\delta_i{}^j)
       \epsilon\,{{\Pi^{-1}}_j}^A S_{\mu\nu\rho ,A}
\eea       
Here $\gamma^\mu$ are the $D=7$ gamma matrices, while $\Gamma^i$ are those
for $SO(5)_c$. Note that $\Gamma^i\lambda_i=0$. Since $\epsilon$ is
a spinor under $SO(5)_c$, the derivative $\nabla_\mu\epsilon$ has both
a spin and an $SO(5)_c$ connection 
\be\label{nabla}
   \nabla_\mu\epsilon = \left( \partial_\mu
       + \frac{1}{4}\omega_\mu{}^{ab}\gamma_{ab}
       + \frac{1}{4}Q_{\mu ij}\Gamma^{ij} \right) \epsilon
\ee

In order to construct dual supersymmetric solutions corresponding
to branes wrapping various supersymmetric cycles, we consider
a metric ansatz of the form
\be\label{metansatz}
ds^2=e^{2 f}[d\xi^2 +d r^2] + e^{2g}\;d\bar{s}^2_\dcy
\ee
Here $d\bar s^2_\dcy$ is the metric on the supersymmetric
$\dcy$-cycle, $\Sigma_d$. We will use $a,b$ to denote tangent space
indices on $\Sigma_d$. The coordinates $\xi^i$, $i=0,\dots 5-\dcy$
span the unwrapped part of the brane with $d\xi^2\equiv \eta_{ij}d\xi^i
d\xi^j=ds^2(\bR^{1,5-\dcy})$.  The functions $f$ and $g$ depend on the radial
coordinate $r$ only. 

The solutions we are interested in have an asymptotic
region with $e^{2f}\approx e^{2g} \approx 1/r^2$, for small $r$,
corresponding to an $AdS_7$-type region with the slices of constant
$r$ given by $\bR^{1,5-\dcy}\times \Sigma_\dcy$, rather than
$\bR^{1,5}$. This asymptotic region is interpreted as specifying the
UV behaviour of the field theory corresponding to the wrapped
fivebrane. The behaviour of the solution in the interior then
specifies the IR behaviour. In all but one case we find an exact
solution of our BPS equations with $g$ constant and $e^{2f}\approx
1/r^2$ corresponding to an $AdS_{(7-\dcy)} \times \Sigma_\dcy$
solution. These solutions are the supergravity duals
of the superconformal theories arising on the wrapped fivebrane.
We will also numerically exhibit flows from the UV $AdS_7$ region
to the $AdS_{(7-\dcy)} \times \Sigma_\dcy$ IR fixed point.

The $SO(5)$-gauge fields for the supergravity solutions 
are specified by the spin connection of the
metric on $\Sigma_\dcy$ corresponding to the fact that the theory
on the M-fivebrane is twisted. In general, we will decompose the $SO(5)$
symmetry into $SO(p)\times SO(q)$ with $p+q=5$, and excite the gauge
fields in the $SO(p)$ subgroup. We will denote these directions by
$m,n=1,\dots,p$. The precise form in each case will be 
given below. Geometrically, in eleven dimensions, the fivebrane is
embedded on a cycle $\Sigma_d$ within a supersymmetric manifold
$M$. This decomposition corresponds to dividing the directions
transverse to the brane into $p$ directions within $M$ and $q$
directions perpendicular to $M$.  
In keeping with this decomposition, the solutions that we
consider will have a single scalar field excited. More precisely we
have 
\be\label{scalaransatz}
{\Pi_A}^i=\diag(e^{q\lambda},\dots,e^{q\lambda},
e^{-p\lambda},\dots,e^{-p\lambda})\ee
where we have $p$ followed by $q$ entries.
Note that this implies that the composite gauge-field $Q$ is then
determined by the gauge-fields via $Q^{ij}=2m B^{ij}$.

For the SLAG five-cycle and most of the four-cycle cases 
the three-form $S$ is non-vanishing. The $S$-equation of motion is 
\be\label{seom}
m^2\delta_{AC}{{\Pi^{-1}}_i}^C{{\Pi^{-1}}_i}^BS_{B}
=-m*F_A+{1\over 4\sqrt 3}\epsilon_{ABCDE}*(F^{BC}\wedge F^{DE})
\ee
and we note that our solutions will have vanishing four-form field
strength $F_A$.  

By substituting this kind of ansatz into the supersymmetry variations 
\p{susytran} and imposing appropriate projections on the spinor
parameters we will then deduce the BPS equations. In the derivation
one finds that it is necessary to twist the gauge connection by the
spin connection, so that
\be\label{conds}
({\bar\omega}^{bc}\gamma_{bc}+2m {B}^{mn}\Gamma_{mn})\epsilon=0
\ee
where ${\bar\omega}^{bc}$ is the spin connection one-form of the
cycle. Essentially, this is in order to set to zero in \p{susytran}
the covariant derivative \p{nabla} in the cycle directions.
After imposing the projections on
$\epsilon$ we are led to identify the appropriate part of the spin
connection of the cycle with the appropriate $SO(5)$ gauge-fields. In
other words, the twisting is dictated by the projections defining
the preserved supersymmetry.

In all cases one finds that, in order to satisfy the BPS equations,
one has the conditions
\bea\label{condtwo}
\gamma^{b}\Gamma_{mn} F^{mn}_{ab}\epsilon&=&
{{\bar R} \over \dcy m} e^{-2g} \;\gamma_a \epsilon \nn
\gamma^{ab}\Gamma_{n} F^{mn}_{ab}\epsilon&=&
{{\bar R} \over pm} e^{-2g} \;\Gamma^m \epsilon
\eea
where $\bar{R}$ is a constant. Using the relation \p{conds} it is easy
to show, from the first condition, that the metric on the cycle is
necessarily Einstein: 
\be\label{ein}
\bar R_{ab}=l \bar g_{ab}
\ee
and so the constant $\bar{R}$ in \p{condtwo} is precisely the Ricci
scalar $\bar{R}=ld$. Given the factor of $e^{2g}$ in \p{metansatz}, we
can rescale $\bar{g}_{ab}$ so that $l=0,\pm 1$. Recall that for $d>
3$ the Einstein condition implies that the Riemann tensor can be written
\be\label{decom}
\bar R_{abcd}=\bar C_{abcd} +{2l\over \dcy -1}\bar g_{a[c}\bar g_{d]b}
\ee
where ${\bar C}$ is the Weyl tensor. For the examples studied
previously, the cycles have been two- or three-dimensional and hence
the Einstein condition implies constant  
curvature i.e. the Reimann  tensor is given by \p{decom}
with $\bar C=0$. For the four- and five-cycles it is
only necessary that the part of the spin
connection involved in the gauging have constant curvature.
We will return to this point and it will be useful to 
refer the Einstein equations which we record here:
\bea\label{einst}
R_{\mu\nu}&=&P_{\mu}P_{\nu}+
(\Pi \Pi F)_{\mu\rho}(\Pi \Pi F)^{\,\,\rho}_\nu
+3m^2({\Pi^{-1}}S)_{\mu\rho\sigma}
(\Pi^{-1} S)_{\nu}{}^{\rho\sigma}\nn
&-&{1\over 10}g_{\mu\nu}\left[m^2(T^2-2T_{ij}T^{ij})
+({\Pi}{\Pi} F)^2
+4m^2(\Pi^{-1}S)^2\right]
\eea  
where contractions over $SO(5)_c$, $SO(5)_g$ and spacetime indices are
implicit.

\section{Special Lagrangian cycles}

Let us first consider fivebranes wrapping special
Lagrangian (SLAG) 3-, 4- and 5-cycles in Calabi--Yau 3-, 4- and 5-folds,
respectively. The dimension $p$ of the transverse space to the fivebrane
within the Calabi--Yau manifold is the same as the dimension of the
cycle $d$. Thus both the holonomy group and the structure 
group of the normal bundle of SLAG $d$-cycles are $SO(d)$. 
The appropriate twisting for such wrappings is obtained by simply
identifying the whole of the $SO(d)$ spin connection  with an $SO(d)$
part of the $R$-symmetry via the splitting $SO(5)\to SO(d)\times
SO(5-d)$.  

This twisting can be seen explicitly by considering
the supersymmetry preserved by fivebranes wrapping the $\dcy$-cycles.
The relevant projections in $D=11$ were written down, for example, in 
section 4.2 of \cite{glw}. In the language of gauged supergravity
we thus impose (in tangent frame)
\bea\label{slag}
\gamma^r\epsilon&=&\epsilon\nn
\gamma^{ab}\epsilon&=&-\Gamma^{ab}\epsilon
\eea
where $a,b=1,\dots,\dcy$ are labelling the directions on the cycle,
The first condition, which is present in all cases, projects the
supersymmetry onto a definite helicity on the fivebrane. The second
conditions describe the twisting, implying that, to satisfy the general
condition \p{conds} that arises in deriving the BPS equations, one
simply sets
\be\label{twist}
\bar\omega_{ab}=2mB_{ab}
\ee
where $B_{ab}$ generate $SO(p)\subset SO(5)$ and we set all
other gauge-fields to zero. Similarly
using the projections in the condition \p{condtwo} one can see
explicitly that the metric on the cycle is indeed Einstein \p{ein}.

Let us now discuss each case further in turn.

\subsection{SLAG three-cycles}
The supersymmetry preserved by a fivebrane wrapping a SLAG three-cycle
corresponds to $N=2$ supersymmetry in $D=3$. Indeed after decoupling gravity,
and considering scales much smaller than the inverse size of the cycle
we obtain an $N=2$ supersymmetric field theory in $D=3$. 

The ansatz for the supergravity BPS solutions 
is given as follows. The metric is given by \p{metansatz} with $\dcy=3$
where the metric on the three-cycle is Einstein. In three dimensions this
implies that it has constant curvature. The scalars are given by
\p{scalaransatz} with $p=3$, $q=2$:
\be\label{slagthreescalars}
{\Pi_A}^i=(e^{2\lambda},e^{2\lambda},e^{2\lambda},
         e^{-3\lambda},e^{-3\lambda})
\ee
The only non-vanishing gauge fields are taken to be $B^{ab}$, for $a,b=1,2,3$,
and these generate $SO(3)\subset SO(5)$. The projections then imply the
twisting \p{twist}. The three-form equation of motion \p{seom}
is solved by setting $S_A=0$.

The resulting BPS equations are given by
\bea
e^{-f}f'&=& - {m\over 10}\left[3e^{-4\lambda}+2e^{6\lambda}\right]
+{3l \over 20m }{e^{4 \lambda-2g}}\nn
e^{-f}{g'} &=& - {m\over 10}\left[3e^{-4\lambda}+2e^{6 \lambda}\right]
- {7l \over 20m }{e^{4 \lambda-2g}}\nn
e^{-f}\lambda' &= & {m\over 5}\left[ 
e^{6\lambda} - e^{-4\lambda}\right] +{l\over 10m } {e^{4\lambda-2g}}
\eea
It should be noted that in this example and for all the cases to 
be considered in this paper, the preserved supersymmetry parameters are
independent of all coordinates except for their radial dependence which is
simply determined by $\delta\psi_r$. In all cases, one finds the
simple dependence $\epsilon=e^{f/2}\epsilon_0$ where $\epsilon_0$ is
constant. Since the Killing spinors are independent of the coordinates
on the cycle we can take arbitrary quotients of the cycle, while
preserving supersymmery. 

When the curvature of the three-cycle is negative, $l=-1$, 
corresponding to a possible quotient of hyperbolic three-space, 
these equations admit a 
solution of the form $AdS_4\times \bH_3$. Specifically we have
\bea\label{stscft}
e^{10\lambda}&=&2\nn
e^{2g}&=&{e^{8\lambda}\over 2m^2}\nn
e^f&=&{e^{4\lambda}\over m}{1\over r}
\eea
In fact this solution was first constructed in \cite{pernicisezgin}.
Here we can interpret it as the dual supergravity 
solution corresponding to the superconformal field theory arising
when an M-fivebrane wraps a SLAG three-cycle $\bH_3$, or a quotient thereof. 
We will analyse the BPS equations numerically in section 4. We will see
there that there are solutions with an $AdS_7$ region for small $r$ describing 
the UV physics of the wrapped brane, which flow to large $r$ corresponding
to the IR physics. We will exhibit a specific flow to the superconformal
fixed point \p{stscft}.

Using the results of \cite{vn,vntwo} we can uplift solutions to the BPS
equations to give supersymmetric solutions in $D=11$. The metric is given by
\be\label{slagthreeup}
ds^2_{11}=\Delta^{-{2\over 5}}ds^2_{7} +{1\over m^2}\Delta^{4\over 5}
\left[e^{4\lambda}DY^aDY^a+e^{-6\lambda}dY^idY^i\right]
\ee
where
\bea\label{slagthreeuptwo}
DY^a&=&dY^a+2mB^{ab}Y^b\nn
\Delta^{-{6\over 5}}&=&e^{-4\lambda}Y^aY^a+e^{6\lambda}Y^iY^i
\eea
where $a=1,2,3$, $i=4,5$ and $(Y^a,Y^i)$ are constrained coordinates
on $\bS^4$ satisfying $Y^aY^a+Y^iY^i=1$. The expression for the four-form
can be found in \cite{vn,vntwo}.

\subsection{SLAG four-cycles}
A fivebrane wrapping a SLAG four-cycle gives rise to $(1,1)$ 
supersymmetry in $D=2$. 
The metric is given by \p{metansatz} with $p=4$, $q=1$ 
and an Einstein metric on the cycle. 
From \p{scalaransatz} the scalars are now given by 
\be\label{slagfourscalars}
{\Pi_A}^i=(e^{\lambda},e^{\lambda},e^{\lambda},
         e^{\lambda},e^{-4\lambda})
\ee
The only non-vanishing gauge fields are taken to be $B^{ab}$, for 
$a,b=1,\dots, 4$, and these generate $SO(4)\subset SO(5)$. 
The projections then imply the twisting \p{twist}.

A new feature for this case is that it is now necessary to switch on the 
three-form $S$. We let
\be\label{S4def}
S_5=-{c\; e^{-8\lambda-4g}\over 64 {\sqrt 3} m^4} \, e^0\wedge e^1
\wedge e^r
\ee
where
\bea
c&=&4m^2e^{4g}\epsilon_{a_1a_2a_3a_4}
\epsilon^{b_1b_2b_3b_4}F^{a_1a_2}_{b_1b_2}F^{a_3a_4}_{b_3b_4}\nn
&=&\epsilon^{a_1a_2a_3a_4}
\epsilon^{b_1b_2b_3b_4}\bar R_{a_1a_2b_1b_2}\bar R_{a_3a_4b_3b_4}
\eea
where in the second line we have used the relation \p{twist} between the
gauge field  $B_{ab}$ and the spin-connection $\bar{\omega}_{ab}$.
If $c$ is constant then the four-form $F_5$ vanishes and the
$S$ equation of motion \p{seom} is satisfied. 

In addition one must also satisfy the Einstein and scalar equations of
motion. Our assumption that the metric on $\Sigma_d$ is Einstein
implies that $\bar R_{ab}$ is proportional to $\bar g_{ab}$ in the Einstein
equations \p{einst}. The ansatz for the scalars and the three-forms
imply that all terms in the right hand side of \p{einst} are proportional
to $g_{ab}$ with the possible exception of the terms
quadratic in the field strength of the gauge-fields. 
Since, by \p{twist}, ${F_{ab}}^{cd}$ is proportional to $\bar{R}_{ab}{}^{cd}$,
to ensure that Einstein's equations are satisfied we must
constrain the Riemann tensor on $\Sigma_d$. 
(An equivalent constraint, requiring that no off-diagonal
scalar fields in ${\Pi_A}^i$ are excited, arises from the scalar
equation of motion.) To get a consistent solution, we will require
that the conformal tensor $\bar{C}_{abcd}$ in the decomposition \p{decom}
vanishes, in which case no problematic terms appear. Given the
Einstein condition, this is equivalent to assuming constant curvature,
so $c$ now depends only on the curvature $l$ of the cycle, and is given
by $c=32l^2/3$.  

The resulting BPS equations then have the form
\bea\label{slagfour}
e^{-f}f'&=& - {m\over 10}\left[4e^{-2\lambda}+e^{8\lambda}\right]
+{l \over 5m }{e^{2 \lambda-2g}}
-{l^2\over 10 m^3}{e^{-4\lambda-4g}}\nn
e^{-f}{g'} &=& - {m\over 10}\left[4e^{-2\lambda}+e^{8 \lambda}\right]
- {3l \over 10m }{e^{2 \lambda-2g}}
+{l^2\over 15 m^3}{e^{-4\lambda-4g}}\nn
e^{-f}\lambda' &= & {m\over 5}\left[ 
e^{8\lambda} - e^{-2\lambda}\right] +{l\over 10m } {e^{2\lambda-2g}}
+{l^2\over 30 m^3}{e^{-4\lambda-4g}}
\eea
If we take the cycle to have constant negative curvature, $l=-1$, we
find that the BPS equations admit the $AdS_3\times \bH^4$ solution 
\bea\label{sfscft}
e^{10\lambda}&=&{3\over 2}\nn
e^{2g}&=&{e^{-6\lambda}\over m^2}\nn
e^f&=&{e^{2\lambda}\over m}{1\over r}
\eea

The uplifted metric in $D=11$ is now given by
\be\label{slagfourup}
ds^2_{11}=\Delta^{-{2\over 5}}ds^2_{7} +{1\over m^2}\Delta^{4\over 5}
\left[e^{2\lambda}DY^aDY^a+e^{-8\lambda}dY^5dY^5\right]
\ee
where 
\bea\label{slagfouruptwo}
DY^a&=&dY^a+2mB^{ab}Y^b\nn
\Delta^{-{6\over 5}}&=&e^{-2\lambda}Y^aY^a+e^{8\lambda}Y^5Y^5
\eea
where $a=1,2,3,4$ with $Y^aY^a+Y^5Y^5=5$. The expression for the four-form
can be found in \cite{vn,vntwo}.

\subsection{SLAG five-cycles}
A fivebrane wrapping a SLAG five-cycle preserves just one supersymmetry.
After decoupling gravity, at low energies we get a quantum mechanical 
model in $D=1$. For this case $d=p=5$ and all of the $SO(5)$
gauge-fields are active, but our ansatz \p{scalaransatz} implies that
all of the scalars to zero:  
\be
{\Pi_A}^i={\delta_A}^i
\ee
All five three-forms are now active and we have
\be
S_a=-{c\; e^{-4g} \over 64{\sqrt 3}m^4}\, e^0\wedge e^r\wedge e^a
\ee
where, given the identification \p{twist} of gauge and spin
connections, 
\bea
c &=& {96\over 5} m^2e^{4g}
       F^{\quad [a_1a_2}_{a_1a_2}F^{\quad a_3a_4]}_{a_3a_4} \\
  &=& {24\over 5}\bar R^{\quad [a_1a_2}_{a_1a_2}
       \bar R^{\quad a_3a_4]}_{a_3a_4}
\eea
To satisfy the $S_A$ equation of motion, we require $c$ to be
constant. As for the four-cycle, this condition and the Einstein's
equations \p{einst} are satisfied if we set $\bar{C}=0$ and take the
five-cycle to have constant curvature, in which case we have $c=6l^2$. 

The BPS equations are given by 
\bea\label{sfibps}
e^{-f}f'&=& - {m\over 2} +{l\over 4 m }e^{-2g} -{9l^2\over 32 m^3}e^{-4g}\nn
e^{-f}{g'} &=& - {m\over 2}
- {l \over 4m }e^{-2g}+{3l^2\over 32 m^3}e^{-4g}
\eea
If we set  $l=-1$ we find the $AdS_2\times \bH_5$ solution
\bea\label{sfiscft}
e^{2g}&=&{3\over 4 m^2}\nn
e^f&=&{3\over 4m}{1\over r}
\eea
On the other hand if we set $l=1$ we find the $AdS_2\times \bS^5$ solution
\bea\label{sfiscftwo}
e^{2g}&=&{1\over 4 m^2}\nn
e^f&=&{1\over 4m}{1\over r}
\eea
The general solution of the BPS equations is presented in section 6.4.

Since the scalars are set to zero, the uplifted $D=11$ metric takes
the simple form
\be
ds^2_{11}=ds^2_{7} +{1\over m^2}DY^aDY^a
\ee
where 
\be
DY^a=dY^a+2mB^{ab}Y^b
\ee
with $Y^aY^a=1$. The expression for the four-form
can be found in \cite{vn,vntwo}.

\section{K\"{a}hler four-cycles}

The spin connection of a K\"{a}hler-cycle is a
$U(2)\approx U(1)\times SU(2)$ connection. The appropriate 
twisting for a fivebrane wrapping a K\"{a}hler cycle
is to identify the $U(1)$ subgroup of 
this spin connection with a $U(1)$ subgroup of the 
$SO(5)$ R-symmetry. Which subgroup depends on whether
the four-cycle is inside a Calabi-Yau three-fold or a Calabi-Yau four-fold.
We now consider each case in turn.

\subsection{K\"{a}hler four-cycles in Calabi-Yau three-folds}
In the case that the four-cycle is in a Calabi-Yau three-fold, 
corresponding to $(4,0)$ supersymmetry in $D=2$, there are two
transverse directions to the five-brane within the three-fold, so
$p=2$. Equivalently the normal bundle has $SO(2)=U(1)$ structure group
and hence the appropriate identification is such that we split
$SO(5)\to SO(2)\times SO(3)$ and identify the $U(1)$ part of the spin
connection with $SO(2)$. 

We let $B^{12}$ generate this $SO(2)$ and set all other gauge
fields to zero. The relevant projections on the 
supersymmetry parameters can be written 
\bea
&\gamma^r\epsilon=\epsilon \nonumber &\\
&\gamma^{12}\epsilon=\gamma^{34}\epsilon=\Gamma^{12}\epsilon&
\eea 
in a basis where the non-vanishing components of the K\"{a}hler-form
on the four-cycle are $J_{12}=J_{34}=1$. We then find that \p{conds} implies
that
\be
B^{12}=-{1\over 4m}\bar\omega_{ab}J^{ab}
\ee
where $a,b=1,\dots,4$ 
and hence the field strength is given by the projection of the Riemann
tensor onto the Ricci-form 
${\bar {\cal R}}_{ab}\equiv {1\over 2} \bar R_{abcd}J^{cd}$:
\be
F^{12}=-{1\over 2m}\bar {\cal R}
\ee
Since we have $p=2,q=3$, given \p{scalaransatz}, the scalar fields are
taken to be 
\be
{\Pi_A}^i=(e^{3\lambda},e^{3\lambda},e^{-2\lambda},
         e^{-2\lambda},e^{-2\lambda})
\ee 
and we can set the 3-form $S$ to zero.

The derivation of the BPS equations again implies that the
metric on the K\"{a}hler cycle is Einstein. Note that we then have
$\bar{\cal R}_{ab}=lJ_{ab}$. In this case, no other constraint
is placed on the cycle. One might expect that, as in the SLAG case,
there is a condition coming from the Einstein equations. For 
SLAG cycles, the conformal part of the curvature \p{decom} was required to
vanish. However, since here the gauge fields depend
only on the $U(1)$ part of the curvature on the cycle, and this
has a vanishing conformal tensor, the stress-energy tensor is 
necessarily proportional to $g_{ab}$ and no such condition arises. 

We obtain the BPS equations
\bea
e^{-f}f'&=& - {m\over 10}\left[2e^{-6\lambda}+3e^{4\lambda}\right]
+{l \over 5m }{e^{6 \lambda-2g}}\nn
e^{-f}{g'} &=& - {m\over 10}\left[2e^{-6\lambda}+3e^{4 \lambda}\right]
- {3l \over 10m }{e^{6 \lambda-2g}}\nn
e^{-f}\lambda' &= & {m\over 5}\left[ 
e^{4\lambda} - e^{-6\lambda}\right] +{l\over 5m } {e^{6\lambda-2g}}
\eea
To look for an $AdS_3\times \Sigma_4$ fixed point we set 
$g'=\lambda'=0$, but find that we are driven to $\lambda\to\infty$.
As for all cases we will numerically investigate these equations in
section~6.

The uplifted metric in $D=11$ is now given by
\be\label{dipsy}
ds^2_{11}=\Delta^{-{2\over 5}}ds^2_{7} +{1\over m^2}\Delta^{4\over 5}
\left[e^{6\lambda}DY^aDY^a+e^{-4\lambda}dY^idY^i\right]
\ee
where 
\bea
DY^a&=&dY^a+2mB^{ab}Y^b\nn
\Delta^{-{6\over 5}}&=&e^{-6\lambda}Y^aY^a+e^{4\lambda}Y^iY^i
\eea
where $a=1,2$, $i=3,4,5$ with $Y^aY^a+Y^iY^i=5$. 
The expression for the four-form
can be found in \cite{vn,vntwo}.

\subsection{K\"{a}hler four-cycles in Calabi-Yau four-folds}

When the K\"{a}hler four-cycle is in a Calabi-Yau four-fold, corresponding
to $(2,0)$ supersymmetry in $D=2$, there are now four directions
transverse to the fivebrane within the four-fold, so
$p=4$. Equivalently, the normal bundle has $U(2)\subset SO(4)$ structure
group. In this case the appropriate identification of the $U(1)$ part 
of the $U(2)$ spin connection is to break $SO(5)\to SO(4)\to U(2)$ and then 
identify the $U(1)$ part of the spin connection with the $U(1)$ in
$U(2)\approx U(1)\times SU(2)$. 

Consequently we take only the $U(1)\subset U(2)$ gauge fields
to be non-vanishing: equivalently we take $B^{12}=B^{34}$ with
all other components vanishing. We have the projections
\bea
&\gamma^r\epsilon=\epsilon, \nonumber &\\
&\gamma^{12}\epsilon=\gamma^{34}\epsilon=\Gamma^{12}\epsilon =
   \Gamma^{34}\epsilon&
\eea
corresponding to the obvious non-vanishing components
of the K\"{a}hler form.
 
We then find
\be
B^{12}+B^{34}=-{1\over 4m}\bar\omega_{ab}J^{ab}
\ee
where $a,b=1,\dots,4$, and hence 
\be
F^{12}+F^{34}=-{1\over 2m}\bar {\cal R}
\ee
In this case, since $p=4,q=1$, the ansatz for the scalars is as in the
SLAG four-cycle case \p{slagfourscalars}.
The ansatz for the 3-form is again as for the SLAG four-cycle case
\p{S4def} but now with
\bea
c&=&4m^2e^{4g}\epsilon_{a_1a_2a_3a_4}
\epsilon^{b_1b_2b_3b_4}F^{a_1a_2}_{b_1b_2}F^{a_3a_4}_{b_3b_4}\nn
&=&16 l^2
\eea
where in the second line we have substituted for $F_{ab}^{cd}$ in
terms of ${\cal R}_{ab}$. As in the previous K\"{a}hler case, we do not
need to impose any additional constraints on the K\"{a}hler-Einstein
metric on the four-cycle. 

The resulting BPS equations then have the form
\bea
e^{-f}f'&=& - {m\over 10}\left[4e^{-2\lambda}+e^{8\lambda}\right]
+{l \over 5m }{e^{2 \lambda-2g}}
-{3l^2\over 20 m^3}{e^{-4\lambda-4g}}\nn
e^{-f}{g'} &=& - {m\over 10}\left[4e^{-2\lambda}+e^{8 \lambda}\right]
- {3l \over 10m }{e^{2 \lambda-2g}}
+{l^2\over 10 m^3}{e^{-4\lambda-4g}}\nn
e^{-f}\lambda' &= & {m\over 5}\left[ 
e^{8\lambda} - e^{-2\lambda}\right] +{l\over 10m } {e^{2\lambda-2g}}
+{l^2\over 20 m^3}{e^{-4\lambda-4g}}
\eea

If we take the cycle to have constant negative curvature, $l=-1$,  
we find the $AdS_3\times \Sigma_4$ solution
\bea\label{kascft}
e^{10\lambda}&=&{4\over 3}\nn
e^{2g}&=&{e^{-6\lambda}\over m^2}\nn
e^f&=&{e^{2\lambda}\over m}{1\over r}
\eea
Note the form of the uplifted metric in $D=11$ is the same as for
the SLAG four-cycles \p{slagfourup}, \p{slagfouruptwo}.

\section{Exceptional cycles}
There are three exceptional calibrations: the associative three-cycles
and the co-associative four-cycles in manifolds of $G_2$-holonomy and
the Cayley four-cycles in manifolds of $Spin(7)$ holonomy. The
supergravity duals of fivebranes wrapping associative three-cycles
was considered in \cite{agk} and here we will analyse the remaining 
two cases.

\subsection{Co-associative four-cycles}
In this case the four-cycle has an $SO(4)\approx SU(2)_L\times SU(2)_R$
spin connection. Since $p=3$, we split the $R$-symmetry $SO(5)\to
SO(3)\times SO(2)$ and the appropriate twisting is obtained by identifying
$SU(2)_L$ with $SO(3)$. This twist leads to (2,0) supersymmetry in $D=2$.

A discussion of the appropriate projections can be found in section
4.3 of \cite{glw}. Here we write these as
\bea
\gamma^r\epsilon&=&\epsilon\nn
\gamma^+_{ab}\epsilon&=&0\nn
\Gamma^{23}\epsilon=\gamma^-_{12}\epsilon\qquad
\Gamma^{31}\epsilon&=&\gamma^-_{13}\epsilon\qquad
\Gamma^{12}\epsilon=\gamma^-_{14}\epsilon
\eea
where the pluses and minuses refer to self-dual and anti-self dual
parts, respectively, and $a,b=1,\dots,4$. 
The $SO(3)$ gauge-fields are generated
by $B^{mn}$, $m,n=1,2,3$ and we set all other gauge-fields to
zero. From \p{conds} we deduce
\bea
\bar\omega^{-12}&=&-mB^{23}\nn
\bar\omega^{-13}&=&-mB^{31}\nn
\bar\omega^{-14}&=&-mB^{12}
\eea

Given $p=3,q=2$, the scalar ansatz is the same as for
the SLAG three-cycles \p{slagthreescalars} and the three-form $S$
can be set to zero.
The condition \p{condtwo} again implies that the 
metric on the cycle is Einstein. In order to ensure Einstein's equations are
solved we note that since, unlike the SLAG case, only the anti-self-dual
part of the spin connection on the cycle enters, it is only necessary
to set $\bar C^-=0$ in \p{decom}. 
In other words we take the associative four-cycle to
have a conformally half-flat Einstein metric. The only compact examples
with $l=1$ are $\bS^4$ or $\bC \bP^2$ and for $l=0$ we have flat space
or $\bK3$. 

The BPS equations are now
\bea
e^{-f}f'&=& - {m\over 10}\left[3e^{-4\lambda}+2e^{6\lambda}\right]
+{l \over 5m }{e^{4 \lambda-2g}}\nn
e^{-f}{g'} &=& - {m\over 10}\left[3e^{-4\lambda}+2e^{6 \lambda}\right]
- {3l \over 10m }{e^{4 \lambda-2g}}\nn
e^{-f}\lambda' &= & {m\over 5}\left[ 
e^{6\lambda} - e^{-4\lambda}\right] +{2l\over 15m } {e^{4\lambda-2g}}
\eea

Setting $l=-1$ we find an $AdS_3\times \Sigma_4$ solution:
\bea\label{cfscft}
e^{10\lambda}&=&3\nn
e^{2g}&=&{e^{8\lambda}\over 3m^2}\nn
e^f&=&{2e^{4\lambda}\over 3m}{1\over r}
\eea

The uplifted solutions in $D=11$ have the same structure as the SLAG
three-cycles \p{slagthreeup},\p{slagthreeuptwo} .

\subsection{Cayley Four-cycles}
The four-cycle has an $SO(4)\approx SU(2)_L\times SU(2)_R$
spin connection. Given now $p=4$, we  split the $R$-symmetry $SO(5)\to
SO(4) \approx SU(2)_L'\times SU(2)_R'$ and the appropriate twisting 
is obtained by identifying $SU(2)_L$ with
$SU(2)_L'$. This twist leads to (1,0) supersymmetry in $D=2$.

Again an explicit discussion of the projections can be found in
section 4.3 of \cite{glw}. Here we will use
\bea
\gamma^r\epsilon&=&\epsilon\nn
\gamma^+_{ab}\epsilon&=&
\Gamma^+_{ab}\epsilon=0\nn
\Gamma^{-}_{ab}\epsilon&=&-\gamma^-_{ab}\epsilon
\eea
The $SU(2)_L'$ gauge-fields are generated
by $B^{-ab}$, $a,b=1,\dots,4$ and we set all other gauge-fields to
zero. From \p{conds} we deduce
\be
\bar\omega^{-ab}=2mB^{-ab}
\ee
Since $p=4,q=1$ the scalar field ansatz is the same as the
SLAG four-cycles and the K\"{a}hler four-cycles in Calabi-Yau four-folds
\p{slagfourscalars}. The three-form $S$ also has the same form \p{S4def},
though now, 
\bea
c&=&4m^2e^{4g}\epsilon_{a_1a_2a_3a_4}
\epsilon^{b_1b_2b_3b_4}F^{a_1a_2}_{b_1b_2}F^{a_3a_4}_{b_3b_4}\nn
&=&4\bar R^-_{abcd}\bar R^{-abcd}\eea
As before, if $c$ is constant then the $S$ equation of motion is
satisfied. 
As in the co-associative case, this condition is satisfied as
are the Einstein equations if we take the cycle to be conformally
half-flat by setting $\bar C^-=0$ in \p{decom}. 
We then get $c=16l^2/3$.

The BPS equations then read
\bea
e^{-f}f'&=& - {m\over 10}\left[4e^{-2\lambda}+e^{8\lambda}\right]
+{l \over 5m }{e^{2 \lambda-2g}}
-{l^2\over 20 m^3}{e^{-4\lambda-4g}}\nn
e^{-f}{g'} &=& - {m\over 10}\left[4e^{-2\lambda}+e^{8 \lambda}\right]
- {3l \over 10m }{e^{2 \lambda-2g}}
+{l^2\over 30 m^3}{e^{-4\lambda-4g}}\nn
e^{-f}\lambda' &= & {m\over 5}\left[ 
e^{8\lambda} - e^{-2\lambda}\right] +{l\over 10m } {e^{2\lambda-2g}}
+{l^2\over 60 m^3}{e^{-4\lambda-4g}}
\eea

If we set $l=-1$ we find the following $AdS_3\times \Sigma_4$ solution
\bea\label{cafscft}
e^{10\lambda}&=&{12\over 7}\nn
e^{2g}&=&{e^{-6\lambda}\over m^2}\nn
e^f&=&{e^{2\lambda}\over m}{1\over r}
\eea

The structure of the uplifted metric in $D=11$ follows the SLAG
four-cycle case and is given by \p{slagfourup} and \p{slagfouruptwo}.

\section{Analysing the BPS equations}
To further analyse the BPS equations it is useful to group
them via the co-dimension of the cycle. 

\subsection{Co-dimension two}
The only co-dimension two-cycle that we have been considering
is the K\"ahler four-cycle in a Calabi-Yau 3-fold. 
Let us introduce the new variables
\bea
\newa^2&=&e^{2g}e^{-12\lambda}\nn
e^\newf&=&e^{f-6\lambda}
\eea
The BPS equations are then somewhat simpler
\bea
e^{-\newf}\newf'&=& - {m\over 2}\left[3e^{10\lambda}-2\right]
-{l\over m\newa^2} \nn
e^{-\newf}{\newa'\over \newa} &=& - {m\over 2}\left[3e^{10\lambda}-2\right]
- {3l\over 2m\newa^2}\nn
e^{-\newf}\lambda' &= & {m\over 5}\left[ 
e^{10\lambda} - 1\right] +{l\over 5m\newa^2}
\eea
The analysis is further simplified by introducing $x=\newa^2$ and
$F=x^{2/3}e^{10\lambda}$, giving the ODE
\be\label{po}
{dF\over dx}={2m^2F\over [3m^2(3Fx^{1\over 3}-2x)+9l]}
\ee
The typical flows in the $(F,x)$-plane for the case of $l=-1$ and
$l=1$ are plotted in figures 1 and 2, respectively.

\begin{figure}[!h]
\vspace{5mm}
\begin{center}
\epsfig{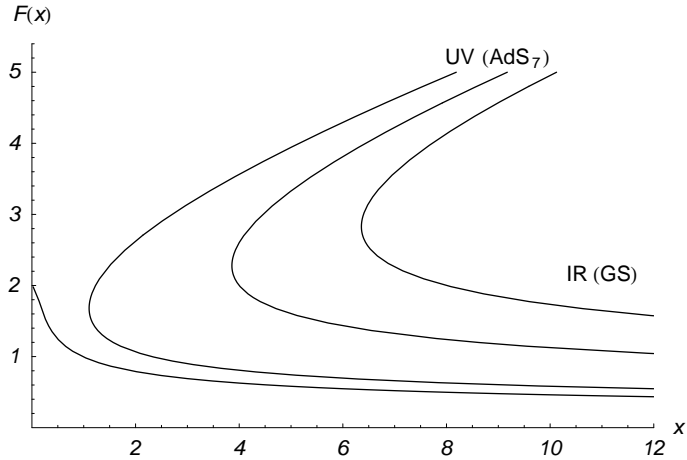}\\
\end{center}
\caption{Behaviour of the orbits for co-dimension two with $l=-1$.
The $AdS_7$-type UV region is when $F$ and $x$ are both large.
The singularity, IR(GS), in the IR region is of the good type.}
\label{fig1}
\vspace{5mm}
\end{figure}  

\begin{figure}[!h]
\vspace{5mm}
\begin{center}
\epsfig{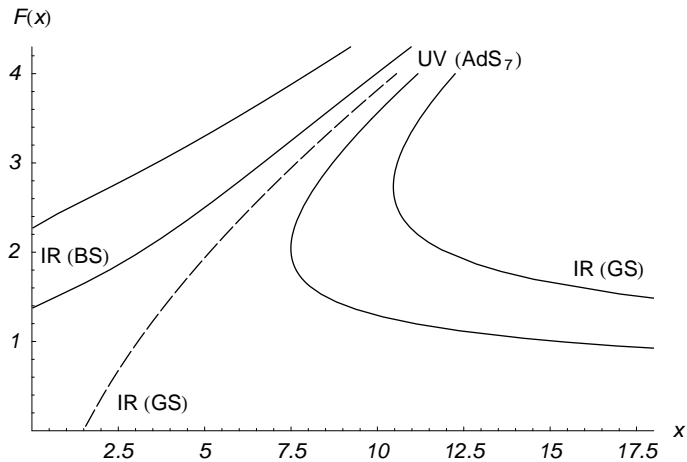}\\
\end{center}
\caption{Behaviour of the orbits for co-dimension two with $l=1$.
IR(GS) and IR(BS) indicate the good and bad singularities in the IR region, 
respectively.}
\label{fig2}
\vspace{5mm}
\end{figure}  

When both $F$ and $x$ are large we get 
$F\approx x^{2\over 3}(1 -2l/m^2x)$. 
Using $\newa$ as a radial variable, we find that
this gives rise to the asymptotic behaviour
\bea
ds^2&=&{4\over m^2\newa^2}d\newa^2 +\newa^2(d\xi^2+d\bar{s}^2_4)\nn
e^{10\lambda}&=&1-{2l\over m^2 \newa^2}.
\eea
This is precisely what we expect for the wrapped M-fivebrane.
The scalars vanish and the metric has the form of $AdS_7$ except that
the slices of constant $\newa$ have $\bR^{1,5}$ replaced with 
$\bR^{1,1}\times\Sigma_4$, where $\Sigma_4$ is the four-cycle with
a K\"ahler-Einstein metric. Note that the next to leading order
behavior of the scalar field corresponds to the insertion
of the boundary operator ${\cal O}_4$ of conformal dimension 
$\Delta=4$ that is dual to an operator constructed from
the scalar fields in the M-fivebrane theory.

The IR behaviour of the wrapped M-fivebrane is obtained by analysing
the asymptotic behaviour of the flows. This case is the exception in
that there is not a flow to an IR $AdS_3\times \Sigma_4$ 
fixed point when $l=-1$. In fact, as one can see from figure
\ref{fig1}, the flows end up in a region of small $F$ and large
$x$. This limit can be analysed explicitly. One finds $F\approx
1/x^{1/3}$ with $e^{10\lambda}\approx 1/x$ tending to zero. The
asymptotic metric is singular and given by 
\be\label{here}
ds^2={1\over m^2\newa^{22\over 5}}
d\newa^2 +\newa^{-{2\over 5}}(d\xi^2+d\bar{s}^2_\dcy)
\ee
It is straightforward to demonstrate the $(00)$ component of
the uplifted $D=11$ metric \p{dipsy} is bounded as we approach the
singularity and hence this is a ``good'' singularity
by the criteria of \cite{malnun}.

For $l=1$, one still has the $AdS_7$-type region at large $F$ and $x$,
but now the flows are different. As can be seen in figure~\ref{fig2},
there are three possibilities. One can flow to the small $F$ and
large $x$ region and one obtains the asymptotic behaviour \p{here} with
a good singularity. A good singularity is also found for the special
orbit with $F=0$ and $x=3l/2m^2$. 
There are also flows to $F$ constant 
and $x=0$ which give rise to bad singularities.

Finally, it is probably worth noting that we can in fact integrate
\p{po} to explicitly realize the behaviour discussed above. In the
original variables one gets the general relation  
\be
-{2l^2\over m^4}\ln(m^2e^{2g-2\lambda}+l)+{2l\over m^2}e^{2g-2\lambda}
-e^{4g-4\lambda}+e^{4g+6\lambda}=C
\ee
for some constant $C$.

\subsection{Co-dimension three}
There are two examples with co-dimension three: the SLAG three-cycles
and the co-associative four-cycles. In this case it is useful
to introduce the new variables
\bea
\newa^2&=&e^{2g}e^{-8\lambda}\nn
e^\newf&=&e^{f-4\lambda}
\eea
The BPS equations are then given by
\bea
e^{-\newf}\newf'&=& - {m\over 2}\left[2e^{10\lambda}-1\right]
-{\gamma\over m\newa^2} \nn
e^{-\newf}{\newa'\over \newa} &=& - {m\over 2}\left[2e^{10\lambda}-1\right]
- {\beta\over m\newa^2}\nn
e^{-\newf}\lambda' &= & {m\over 5}\left[ 
e^{10\lambda} - 1\right] +{\alpha\over 2m\newa^2}
\eea
where $(\alpha,\beta,\gamma)$=$(l/5,3l/4,l/4)$
for the SLAG three-cycles and $(4l/15,5l/6,l/3)$
for the associative four-cycles.
We next define $x=\newa^2$ and $F=xe^{10\lambda}$ and obtain the
ODE
\be
{dF\over dx}={F[m^2x-5\alpha+2\beta]\over x[m^2(2F-x)+2\beta]}
\ee

\begin{figure}[!htb]
\vspace{5mm}
\begin{center}
\epsfig{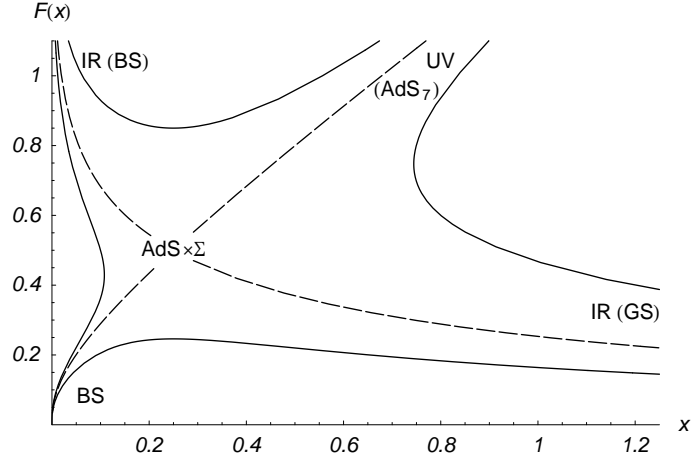}\\
\end{center}
\caption{Behaviour of the orbits for co-dimension three with $l=-1$.
Note the flow from the $AdS_7$-type region when $F,x$ are large to
the IR fixed point and the flows to the good and bad singularities in the IR,
IR(GS) and IR(BS), respectively.}
\label{fig3}
\vspace{5mm}
\end{figure}   
\begin{figure}[!htb]
\vspace{5mm}
\begin{center}
\epsfig{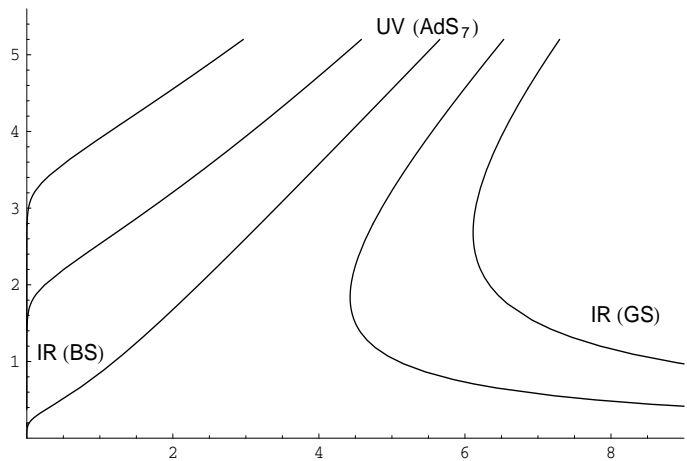}\\
\end{center}
\caption{Behaviour of the orbits for co-dimension three with $l=1$.}
\label{fig4}
\vspace{5mm}
\end{figure}  

The typical behaviour of $F(x)$ is illustrated in figures 3 for $l=-1$
and figure 4 for $l=1$. The region where both
$x$ and $F$ large corresponds to the
$AdS_7$ type region describing the UV behaviour of the wrapped brane.
We have $F\approx x-5\alpha/m^2$ and using $\newa$ as a radial variable
we obtain the asymptotic behaviour
\bea
ds^2&=&{4\over m^2\newa^2}d\newa^2 +\newa^2(d\xi^2+d\bar{s}^2_\dcy)\nn
e^{10\lambda}&=&1-{5\alpha\over m^2 \newa^2}.
\eea
Again we see that the operator ${\cal O}_4$ is switched on.

For $l=-1$ we can flow from the UV region to the $AdS\times \Sigma_d$
fixed point that was given in \p{stscft} and \p{cfscft} for the SLAG
three-cycles and co-associative cycles, respectively. There are also 
flows exhibited in figure 3 which flow to small $F$ for large $x$. These
behave like $F\approx 1/x$ with $e^{10\lambda}\approx 1/x^2$
tending to zero. The asymptotic metric is given by
\be\label{tink}
ds^2={4\over m^2\newa^{26\over 5}}
d\newa^2 +\newa^{-{6\over 5}}(d\xi^2+d\bar{s}^2_\dcy)
\ee
It is straightforward to demonstrate that these are good singularities. 
There are also flows from the $AdS_7$ region
to large $F$ and small $x$. They have $F\approx ((2\beta -5\alpha)/2m^2)\ln x$
and give rise to bad singularities. Similarly the flow from the $AdS_3$ fixed 
point to small $F$ and $x$ have $F\approx x^{(2\beta -5\alpha)/2\beta}$
and give bad singularities.

When $l=1$ the flows from the UV to the IR are illustrated in figure 4.
The flows to small $F$ and large $x$ give rise to the asymptotic behaviour
\p{tink} and hence have good singularities. The singularities for
the flows to small $F$ and $x$ are the same as for $l=-1$ and hence are bad.

\subsection{Co-dimension four}
There are three examples with co-dimension four: SLAG four-cycles,
K\"ahler four-cycles in Calabi-Yau four-folds and Cayley four-cycles.
It is now convenient to introduce the new variables
\bea
\newa^2&=&e^{2g}e^{-4\lambda}\nn
e^\newf&=&e^{f-2\lambda}
\eea
The BPS equations are then given by
\bea
e^{-\newf}\newf'&=& - {m\over 2}e^{10\lambda}
-{\beta \over 2e^{10\lambda}\newa^4} \nn
e^{-\newf}{\newa'\over \newa} &=& - {m\over 2}e^{10\lambda}
- {\alpha\over 2\newa^2}\nn
e^{-\newf}\lambda' &= & {m\over 5}\left[ 
e^{10\lambda} - 1\right] +{\alpha\over 10\newa^2}+{\beta\over 10e^{10\lambda}
\newa^4}
\eea
where $\alpha={l/m}$ and $\beta={l^2/3m^3}$, $l^2/2m^3$ and
$l^2/6m^3$ for the SLAG, K\"ahler and Cayley four-cycles,
respectively. 

We next define $x=\newa^2$ and $F=x^2e^{10\lambda}$ and obtain the
ODE
\be
{dF\over dx}={F(\alpha +2mx)-\beta x\over mF+\alpha x}
\ee
The typical behaviour of $F(x)$ is illustrated in figures 5 for $l=-1$
and figure 6 for $l=1$. The region of $x$ and $F$ large corresponds to the
$AdS_7$-type region describing the UV behaviour of the wrapped brane.
We have $F\approx x^2-(\alpha/m) x$. Using $\newa$ as a radial variable
we obtain the asymptotic behaviour
\bea
ds^2&=&{4\over m^2\newa^2}d\newa^2 +\newa^2(d\xi^2+d\bar{s}^2_\dcy)\nn
e^{10\lambda}&=&1-{\alpha\over m\newa^2}.
\eea
The asymptotic behavior of the scalar again indicates that ${\cal O}_4$
is switched on.

For $l=-1$ we can flow from the UV region to the $AdS\times \Sigma_4$
fixed points that were given in \p{sfscft}, \p{kascft} and \p{cafscft} for
the SLAG, K\"ahler and Cayley four-cycles, respectively. There are 
also flows exhibited in figure 5 which flow to $F=\beta/2m$ for large $x$. 
We then have $e^{10\lambda}\approx (\beta/2m)/x^2$ tending to zero. 
Again it is straightforward to demonstrate that these are good singularities. 
There are also flows from the $AdS_7$ region
to constant $F$ and small $x$. The asymptotic metrics have bad singularities.

The behaviour of the flows for $l=1$ are illustrated in figure 6.
The flows from the UV region end up with $F$ constant when $x=0$
and have bad singularities.

\begin{figure}[!htb]
\vspace{5mm}
\begin{center}
\epsfig{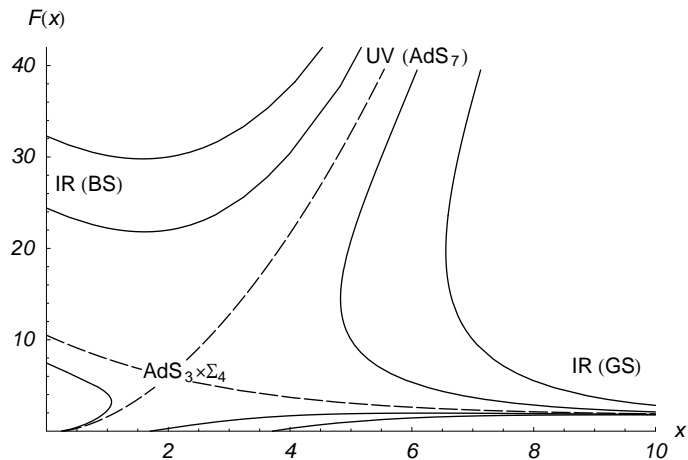}\\
\end{center}
\caption{Behaviour of the orbits for co-dimension four with $l=-1$.
Note the flow from the $AdS_7$-type region when $F,x$ are large to
the IR fixed point and the flows to the good and bad singularities in the IR,
IR(GS) and IR(BS), respectively.}
\label{fig5}
\vspace{5mm}
\end{figure}  
\begin{figure}[!htb]
\vspace{5mm}
\begin{center}
\epsfig{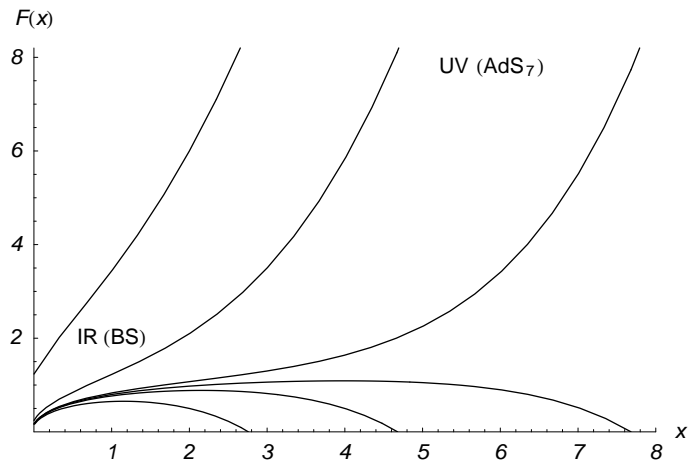}\\
\end{center}
\caption{Behaviour of the orbits for co-dimension four with $l=1$.}
\label{fig6}
\vspace{5mm}
\end{figure}  

We conclude this subsection by determining
the central charges of the two dimensional
conformal field theories arising at the fixed points of the flows
by generalising the argument
of \cite{malnun}. We use 
\be
c={3R_{AdS_3}\over 2 G_3}
\ee
and relate the three-dimensional Newton's constant
$G_{3}$ to the eleven-dimensional Newton's constant
as in \cite{malnun}. To do this we work with units where
the radius of $AdS_7$ in the $AdS_7\times \bS^4$ solution is one
by setting $m=2$. We then find
\be
c={8N^3\over \pi^2}e^{f_0+4g} \textrm{Vol}(\bar\Sigma)
\ee
where $\textrm{Vol}(\bar\Sigma)$ is the volume of the four-cycle
and $e^f\equiv e^{f_0}/r$ at the fixed point. From 
\p{sfscft}, \p{kascft}, \p{cfscft} and \p{cafscft}
we get $e^{f_0+4g}=1/48,3/128,1/48$ and $7/384$ for the SLAG, K\"ahler
four-cycles in four-fold, co-associative and Cayley four-cycles,
respectively.      

\subsection{Co-dimension five}

The SLAG five-cycle is the only co-dimension five case. It is rather
different than the other cases in that the scalars are all set to
zero. To solve the BPS equations \p{sfibps} we first introduce
a new radial variable $\rho$ via
\be
{d\rho\over dr}=e^{2f}
\ee
We then find the general solution is given by
\be
ds^2=-e^{2f}dt^2+e^{-2f}d\rho^2 +\rho^2 d\bar s^2_5
\ee
with
\be
e^{2f} = {m^2\over 4 \rho^6}
   \left(\rho^2-{l\over 4m^2}\right)^2
   \left(\rho^2+{3l\over 4m^2}\right)^2
\ee
which flows for $l=-1$ or $l=1$ to the conformal fixed points given
in \p{sfiscft} or \p{sfiscftwo}, respectively.

\section{Discussion}

We have presented a large class of supergravity solutions that are
dual to the twisted theories arising on M-fivebranes wrapping general 
supersymmetric cycles. An Einstein metric on the cycle is an ingredient
in the construction: for the SLAG cycles it must have constant curvature,
for the K\"ahler cycles it must be K\"ahler-Einstein, for the co-associative
and Cayley four-cycles it must be conformally half-flat.

The solutions have an asymptotic $AdS_7$ type-region 
that describes the UV physics. When the curvature of the Einstein metric
on the cycle is negative, $l=-1$, in all but one case, K\"{a}hler four-cycles
in Calabi-Yau three-folds, there is a flow to an IR fixed point of
the form $AdS_{7-\dcy}\times \Sigma_\dcy$. These fixed points are dual
to the superconformal field theories arising on the M-fivebrane and
thus provide new examples of AdS/CFT duality. For positive curvature, $l=1$,
we only found such a fixed point for SLAG five-spheres.
We also exhibited flows to other IR limits and determined 
whether the resulting singularities were of a good or bad type according 
to the criteria of \cite{malnun}. It will be interesting to study all of the 
IR physics in more detail. 
When the cycle is Ricci-flat, $l=0$, the cycle can either be
flat or for  K\"ahler, co-associative or
Cayley four-cycles it can also be $\bK3$ (if we relax compactness it could
be any four manifold with $SU(2)$ holonomy). In this case the gauge fields are 
zero and there is no twisting and so we simply have a fivebrane wrapping 
$\bT^4$ or $\bK3$, whose supergravity solutions are well known.

The solutions that have been constructed here and in \cite{malnun,agk,no} 
have the minimal gauge fields active consistent with the required twisting. 
It would be interesting to generalise our solutions to include
more general gauge-fields which correspond to cycles with the most
general normal-bundles. Note, for example, that this would distinguish
fivebranes wrapping four-cycles in eight-manifolds with $Sp(2)$
holonomy from those with $SU(4)$. It would also be interesting to try
an find solutions that relax the Einstein condition. It is possible
that such generalisations will involve activating more than a single
scalar field. Another direction to pursue is to construct supergravity
solutions corresponding to having both fivebranes and membranes
involved. For example, it might be possible to construct supergravity
solutions analogous to the configurations that were investigated from
the M-fivebrane world-volume point of view in \cite{glwtwo,gaunt}. 

\section{Acknowledgements}
We thank Bobby Acharya and Fay Dowker for discussions.
JPG thanks the EPSRC for partial support, DJW thanks the
Royal Society for support.
All authors are supported in part by PPARC through SPG $\#$613.    


\end{document}